\documentstyle[12pt]{article}
\def\fun#1#2{\lower3.6pt\vbox{\baselineskip0pt\lineskip.9pt
  \ialign{$\mathsurround=0pt#1\hfil##\hfil$\crcr#2\crcr\sim\crcr}}}
\newcommand{\dd}{\mbox{d}}
\newcommand{\Li}{\mbox{Li}_2}

\title{\bf
Compton tensor with heavy photon in the case of
longitudinally polarized fermion.
}

\author{I.~Akushevich$^{1}$, A.~Arbuzov$^{2}$ and E.~Kuraev$^{2}$}

\date{}

\begin{document}

\maketitle

\begin{center}
{
$^{1}$ \it National Center of Particle and
 High Energy Physics, Minsk, 220040, Belarus \\
$^{2}$ \it Joint Institute for Nuclear Research,
Dubna, 141980, Russia
}
\end{center}

\begin{abstract}
The matrix element squared for Compton scattering
of a heavy photon on a longitudinally polarized
electron is calculated.
The gauge--invariant tensor structure of the heavy photon
polarization is considered, assuming that all kinematical
invariants and the invariant mass of the heavy photon are large
compared with the electron mass. The expressions for the Born and
first order radiative corrections are obtained.
Applications are discussed.
\end{abstract}


In the case of unpolarized fermions the Compton tensor with
heavy photon was calculated in papers~\cite{r1,r2} years ago.
It accumulates a considerable part of radiative corrections
and can be used as a building block in calculations of various
processes. The tensor was used for precision calculations of
radiative corrections to Bhabha scattering at LEP~\cite{sabs},
cross--section of deep inelastic scattering with tagged
photons~\cite{tdis}, and other.

We will restrict ourselves here in considering only
that part of the Compton tensor, which contains the degree of
polarization of the initial electron.
The absence in literature of a closed expression for this quantity
and the importance of it for many applications is the
motivation for this investigation.

Let us consider the process (see Fig.~1)
\begin{eqnarray} \label{eq:1}
&& \gamma^*(q)\ +\ e(p_1) \longrightarrow \gamma(k_1)\ +\ e(p_2),
\\ \nonumber
&& q^2 < 0,\quad k_1^2 = 0,\quad p_1^2 = p_2^2 = m^2, \quad
p_1 + q = p_2 + k_1,
\end{eqnarray}
where $m$ is the electron mass.

The Compton tensor is defined as
\begin{eqnarray} \label{eq:2}
K_{\rho\sigma}=(8\pi\alpha)^{-2}\Sigma M_\rho^{\vec{e}\gamma^* \to e\gamma}
(M_\sigma^{{\vec e}\gamma^* \to e\gamma})^*,
\end{eqnarray}
where the matrix element $M$ describes the Compton scattering
process~(\ref{eq:1}). It is read
\begin{eqnarray}
M_\rho &=& M_{0\rho} + M_{1\rho} =
\bar u(p_2)O_{\rho\mu}u(p_1)e_\lambda^\mu(k_1),
\nonumber \\
O_{\rho\mu} &=& O_{\rho\mu}^{(0)} +
\frac{\alpha}{4\pi}O_{\rho\mu}^{(1)}, \quad
O_{\rho\mu}^{(0)} = \gamma_\rho\frac{({\hat p}_2-{\hat q}+m)}{t}\gamma_\mu
+ \gamma_\mu\frac{({\hat p}_1+{\hat q}+m)}{s}\gamma_\rho,
\end{eqnarray}
The quantities $O_{\rho\mu}^{(0)}$ and $O_{\rho\mu}^{(1)}$ take
into account the lowest and the first orders of perturbation theory
respectively.

\vspace{.3cm}
\unitlength=2.10pt
\special{em:linewidth 0.4pt}
\linethickness{0.4pt}
\begin{picture}(165.00,66.33)
\put(80.00,27.00){\makebox(0,0)[cc]{$\times$}}
\put(84.50,32.00){\makebox(0,0)[cc]{$\gamma^*$}}
\put(80.00,28.00){\oval(2.00,2.00)[l]}
\put(80.00,30.00){\oval(2.00,2.00)[r]}
\put(80.00,32.00){\oval(2.00,2.00)[l]}
\put(80.00,34.00){\oval(2.00,2.00)[r]}
\put(80.00,36.00){\oval(2.00,2.00)[l]}
\put(80.00,38.00){\oval(2.00,2.00)[r]}
\put(80.00,40.00){\oval(2.00,2.00)[l]}
\put(80.00,42.00){\oval(2.00,2.00)[r]}
\put(80.00,44.00){\oval(2.00,2.00)[l]}
\put(80.00,45.00){\vector(1,1){10.00}}
\put(80.00,45.00){\line(1,1){20.00}}
\put(80.00,45.00){\line(-1,1){20.00}}
\put(70.00,55.00){\vector(1,-1){5.00}}
\put(60.00,65.00){\vector(1,-1){5.00}}
\put(70.80,55.33){\oval(2.00,3.33)[lt]}
\put(70.80,58.67){\oval(2.00,3.33)[rb]}
\put(72.80,58.33){\oval(2.00,3.33)[lt]}
\put(72.80,61.67){\oval(2.00,3.33)[rb]}
\put(74.80,61.33){\oval(2.00,3.33)[lt]}
\put(74.80,64.67){\oval(2.00,3.33)[rb]}
\put(76.80,64.33){\oval(2.00,3.33)[lt]}
\put(56.14,63.00){\makebox(0,0)[cc]{$\vec{e}$}}
\put(103.00,63.00){\makebox(0,0)[cc]{$e$}}
\put(80.00,63.00){\makebox(0,0)[cc]{$\gamma$}}
\
\put(140.00,27.00){\makebox(0,0)[cc]{$\times$}}
\put(144.00,32.00){\makebox(0,0)[cc]{$q$}}
\put(140.00,28.00){\oval(2.00,2.00)[l]}
\put(140.00,30.00){\oval(2.00,2.00)[r]}
\put(140.00,32.00){\oval(2.00,2.00)[l]}
\put(140.00,34.00){\oval(2.00,2.00)[r]}
\put(140.00,36.00){\oval(2.00,2.00)[l]}
\put(140.00,38.00){\oval(2.00,2.00)[r]}
\put(140.00,40.00){\oval(2.00,2.00)[l]}
\put(140.00,42.00){\oval(2.00,2.00)[r]}
\put(140.00,44.00){\oval(2.00,2.00)[l]}
\put(140.00,45.00){\vector(1,1){5.00}}
\put(140.00,45.00){\vector(1,1){15.00}}
\put(140.00,45.00){\line(1,1){20.00}}
\put(120.00,65.00){\vector(1,-1){10.00}}
\put(120.00,65.00){\line(1,-1){20.00}}
\put(149.80,55.33){\oval(2.00,3.33)[rt]}
\put(149.80,58.67){\oval(2.00,3.33)[lb]}
\put(147.80,58.33){\oval(2.00,3.33)[rt]}
\put(147.80,61.67){\oval(2.00,3.33)[lb]}
\put(145.80,61.33){\oval(2.00,3.33)[rt]}
\put(145.80,64.67){\oval(2.00,3.33)[lb]}
\put(143.80,64.33){\oval(2.00,3.33)[rt]}
\put(116.14,63.00){\makebox(0,0)[cc]{$p_1$}}
\put(164.14,63.00){\makebox(0,0)[cc]{$p_2$}}
\put(141.00,63.00){\makebox(0,0)[cc]{$k_1$}}
\end{picture}
\vspace{-1.3cm}

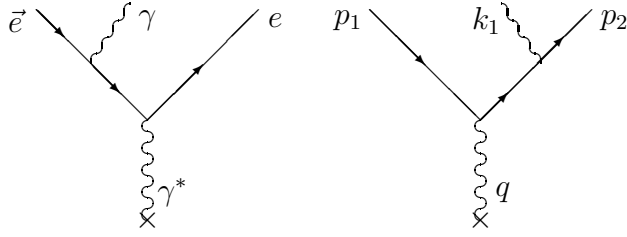
{\small {\bf Fig.~1:} The Born--level Feynman diagrams.}
\vspace{.3cm}

Calculating the first order correction, we will assume that all
kinematical invariants of the process are large in comparison
with the electron mass squared:
\begin{eqnarray}
&& s \sim -t \sim -u \sim -q^2 \gg m^2, \nonumber \\
&& s = 2p_2k_1,\quad t=-2p_1k_1,\quad u=-2p_1p_2,\quad q^2=s+t+u.
\end{eqnarray}
So, we will neglect the electron mass in all places, where possible.
Note that for the unpolarized case in~\cite{r1} the mass was taken
into account.

\begin{figure}[h]
\unitlength=2.10pt
\special{em:linewidth 0.4pt}
\linethickness{0.4pt}
\begin{picture}(223.81,75.33)
\put(143.14,28.00){\oval(2.00,2.00)[l]}
\put(143.14,30.00){\oval(2.00,2.00)[r]}
\put(143.14,32.00){\oval(2.00,2.00)[l]}
\put(143.14,34.00){\oval(2.00,2.00)[r]}
\put(143.14,36.00){\oval(2.00,2.00)[l]}
\put(143.14,38.00){\oval(2.00,2.00)[r]}
\put(143.14,40.00){\oval(2.00,2.00)[l]}
\put(143.14,42.00){\oval(2.00,2.00)[r]}
\put(143.14,44.00){\oval(2.00,2.00)[l]}
\put(123.14,70.00){\line(4,-5){20.00}}
\put(143.14,45.00){\line(4,5){20.00}}
\put(135.14,56.51){\oval(2.00,2.33)[t]}
\put(137.14,56.67){\oval(2.00,2.67)[b]}
\put(139.14,56.51){\oval(2.00,2.33)[t]}
\put(141.14,56.67){\oval(2.00,2.67)[b]}
\put(143.14,56.51){\oval(2.00,2.33)[t]}
\put(145.14,56.67){\oval(2.00,2.67)[b]}
\put(147.14,56.51){\oval(2.00,2.33)[t]}
\put(149.14,56.67){\oval(2.00,2.67)[b]}
\put(151.14,56.51){\oval(2.00,2.33)[t]}
\put(153.81,42.00){\makebox(0,0)[cc]{$(c)$}}
\put(143.14,27.00){\makebox(0,0)[cc]{$\times$}}
\put(125.80,66.67){\vector(3,-4){1.67}}
\put(138.80,50.67){\vector(3,-4){1.67}}
\put(199.81,28.00){\oval(2.00,2.00)[l]}
\put(199.81,30.00){\oval(2.00,2.00)[r]}
\put(199.81,32.00){\oval(2.00,2.00)[l]}
\put(199.81,34.00){\oval(2.00,2.00)[r]}
\put(199.81,36.00){\oval(2.00,2.00)[l]}
\put(199.81,38.00){\oval(2.00,2.00)[r]}
\put(199.81,40.00){\oval(2.00,2.00)[l]}
\put(199.81,42.00){\oval(2.00,2.00)[r]}
\put(199.81,44.00){\oval(2.00,2.00)[l]}
\put(179.81,70.00){\line(4,-5){20.00}}
\put(199.81,45.00){\line(4,5){20.00}}
\put(187.81,61.51){\oval(2.00,2.33)[t]}
\put(189.81,61.67){\oval(2.00,2.67)[b]}
\put(191.81,61.51){\oval(2.00,2.33)[t]}
\put(193.81,61.67){\oval(2.00,2.67)[b]}
\put(195.81,61.51){\oval(2.00,2.33)[t]}
\put(197.81,61.67){\oval(2.00,2.67)[b]}
\put(199.81,61.51){\oval(2.00,2.33)[t]}
\put(201.81,61.67){\oval(2.00,2.67)[b]}
\put(203.81,61.51){\oval(2.00,2.33)[t]}
\put(205.81,61.67){\oval(2.00,2.67)[b]}
\put(207.81,61.51){\oval(2.00,2.33)[t]}
\put(209.81,61.67){\oval(2.00,2.67)[b]}
\put(211.81,61.51){\oval(2.00,2.33)[t]}
\put(194.81,52.66){\oval(2.00,3.33)[lt]}
\put(194.81,56.00){\oval(2.00,3.33)[rb]}
\put(196.81,55.66){\oval(2.00,3.33)[lt]}
\put(196.81,59.00){\oval(2.00,3.33)[rb]}
\put(198.81,58.66){\oval(2.00,3.33)[lt]}
\put(198.81,62.00){\oval(2.00,3.33)[rb]}
\put(200.81,61.66){\oval(2.00,3.33)[lt]}
\put(200.81,65.00){\oval(2.00,3.33)[rb]}
\put(202.81,64.66){\oval(2.00,3.33)[lt]}
\put(202.81,68.00){\oval(2.00,3.33)[rb]}
\put(204.81,67.66){\oval(2.00,3.33)[lt]}
\put(204.81,71.00){\oval(2.00,3.33)[rb]}
\put(199.81,27.00){\makebox(0,0)[cc]{$\times$}}
\put(209.81,42.00){\makebox(0,0)[cc]{$(d)$}}
\
\put(182.47,66.67){\vector(3,-4){1.67}}
\put(189.14,58.67){\vector(3,-4){1.67}}
\put(195.47,50.67){\vector(3,-4){1.67}}
\put(131.80,60.33){\oval(2.00,3.33)[lt]}
\put(131.80,63.67){\oval(2.00,3.33)[rb]}
\put(133.80,63.33){\oval(2.00,3.33)[lt]}
\put(133.80,66.67){\oval(2.00,3.33)[rb]}
\put(135.80,66.33){\oval(2.00,3.33)[lt]}
\put(135.80,69.67){\oval(2.00,3.33)[rb]}
\put(137.80,69.33){\oval(2.00,3.33)[lt]}
\put(130.80,60.67){\vector(3,-4){1.67}}
\put(157.47,62.67){\vector(3,4){1.67}}
\put(146.14,48.67){\vector(3,4){1.67}}
\put(214.14,62.67){\vector(3,4){1.67}}
\put(202.47,48.34){\vector(3,4){1.67}}
\put(120.14,68.00){\makebox(0,0)[cc]{$p_1$}}
\put(175.81,68.00){\makebox(0,0)[cc]{$p_1$}}
\put(167.14,68.00){\makebox(0,0)[cc]{$p_2$}}
\put(223.81,68.00){\makebox(0,0)[cc]{$p_2$}}
\put(146.00,61.00){\makebox(0,0)[cc]{$k$}}
\put(141.14,68.00){\makebox(0,0)[cc]{$k_1$}}
\put(209.14,68.00){\makebox(0,0)[cc]{$k_1$}}
\
\put(34.00,28.00){\oval(2.00,2.00)[l]}
\put(34.00,30.00){\oval(2.00,2.00)[r]}
\put(34.00,32.00){\oval(2.00,2.00)[l]}
\put(34.00,34.00){\oval(2.00,2.00)[r]}
\put(34.00,36.00){\oval(2.00,2.00)[l]}
\put(34.00,38.00){\oval(2.00,2.00)[r]}
\put(34.00,40.00){\oval(2.00,2.00)[l]}
\put(34.00,42.00){\oval(2.00,2.00)[r]}
\put(34.00,44.00){\oval(2.00,2.00)[l]}
\put(14.00,70.00){\line(4,-5){20.00}}
\put(34.00,45.00){\line(4,5){20.00}}
\put(21.00,62.51){\oval(2.00,2.00)[t]}
\put(23.00,62.51){\oval(2.00,2.00)[b]}
\put(25.00,62.51){\oval(2.00,2.00)[t]}
\put(27.00,62.51){\oval(2.00,2.00)[b]}
\put(29.00,62.51){\oval(2.00,2.00)[t]}
\put(31.00,62.51){\oval(2.00,2.00)[lb]}
\put(31.00,60.51){\oval(2.00,2.00)[r]}
\put(31.00,58.51){\oval(2.00,2.00)[l]}
\put(31.00,56.51){\oval(2.00,2.00)[r]}
\put(31.00,54.51){\oval(2.00,2.00)[l]}
\put(31.00,52.51){\oval(2.00,2.00)[r]}
\put(31.00,50.51){\oval(3.00,2.00)[lt]}
\put(16.00,67.50){\vector(3,-4){1.67}}
\put(24.00,57.50){\vector(3,-4){1.67}}
\put(30.00,50.00){\vector(3,-4){1.67}}
\put(43.00,56.25){\vector(3,4){1.67}}
\put(19.80,64.33){\oval(2.00,3.33)[lt]}
\put(19.80,67.67){\oval(2.00,3.33)[rb]}
\put(21.80,67.33){\oval(2.00,3.33)[lt]}
\put(21.80,70.67){\oval(2.00,3.33)[rb]}
\put(23.80,70.33){\oval(2.00,3.33)[lt]}
\put(23.80,73.67){\oval(2.00,3.33)[rb]}
\put(44.81,42.00){\makebox(0,0)[cc]{$(a)$}}
\put(39.81,33.00){\makebox(0,0)[cc]{$\gamma^{*}$}}
\put(34.00,27.00){\makebox(0,0)[cc]{$\times$}}
\put(29.00,71.00){\makebox(0,0)[cc]{$\gamma$}}
\put(10.14,68.00){\makebox(0,0)[cc]{$\vec{e}$}}
\put(57.14,68.00){\makebox(0,0)[cc]{$e$}}
\
\put(89.00,28.00){\oval(2.00,2.00)[l]}
\put(89.00,30.00){\oval(2.00,2.00)[r]}
\put(89.00,32.00){\oval(2.00,2.00)[l]}
\put(89.00,34.00){\oval(2.00,2.00)[r]}
\put(89.00,36.00){\oval(2.00,2.00)[l]}
\put(89.00,38.00){\oval(2.00,2.00)[r]}
\put(89.00,40.00){\oval(2.00,2.00)[l]}
\put(89.00,42.00){\oval(2.00,2.00)[r]}
\put(89.00,44.00){\oval(2.00,2.00)[l]}
\put(69.00,70.00){\line(4,-5){20.00}}
\put(89.00,45.00){\line(4,5){20.00}}
\put(76.00,62.51){\oval(2.00,2.00)[t]}
\put(78.00,62.51){\oval(2.00,2.00)[b]}
\put(80.00,62.51){\oval(2.00,2.00)[t]}
\put(82.00,62.51){\oval(2.00,2.00)[b]}
\put(84.00,62.51){\oval(2.00,2.00)[t]}
\put(86.00,62.51){\oval(2.00,2.00)[lb]}
\put(86.00,60.51){\oval(2.00,2.00)[r]}
\put(86.00,58.51){\oval(2.00,2.00)[l]}
\put(86.00,56.51){\oval(2.00,2.00)[r]}
\put(86.00,54.51){\oval(2.00,2.00)[l]}
\put(86.00,52.51){\oval(2.00,2.00)[r]}
\put(86.00,50.51){\oval(3.00,2.00)[lt]}
\put(71.00,67.50){\vector(3,-4){1.67}}
\put(85.80,49.33){\vector(3,-4){1.67}}
\put(98.00,56.25){\vector(3,4){1.67}}
\put(82.00,55.00){\oval(2.00,3.00)[lt]}
\put(82.00,58.00){\oval(2.00,3.00)[rb]}
\put(84.00,58.00){\oval(2.00,3.00)[lt]}
\put(84.00,61.00){\oval(2.00,3.00)[rb]}
\put(86.00,61.00){\oval(2.00,3.00)[lt]}
\put(86.00,64.00){\oval(2.00,3.00)[rb]}
\put(88.00,64.00){\oval(2.00,3.00)[lt]}
\put(88.00,67.00){\oval(2.00,3.00)[rb]}
\put(90.00,67.00){\oval(2.00,3.00)[lt]}
\put(90.00,70.00){\oval(2.00,3.00)[rb]}
\put(99.81,42.00){\makebox(0,0)[cc]{$(b)$}}
\put(89.00,27.00){\makebox(0,0)[cc]{$\times$}}
\put(86.00,71.00){\makebox(0,0)[cc]{$k_1$}}
\put(65.00,68.00){\makebox(0,0)[cc]{$p_1$}}
\put(112.00,68.00){\makebox(0,0)[cc]{$p_2$}}
\put(94.00,33.00){\makebox(0,0)[cc]{$q$}}
\end{picture}
\end{figure}
\vspace{-1.3cm}
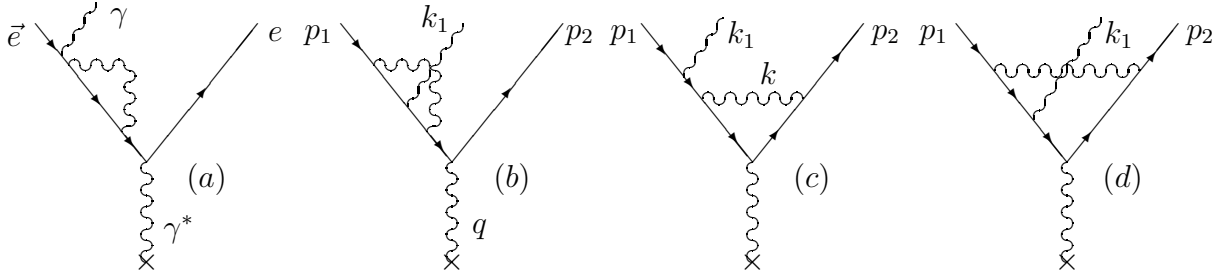
{\small {\bf Fig.~2:} One--loop virtual Feynman diagrams
with photon emission by the {\em initial} electron.}
\vspace{.3cm}

The Compton tensor defined in (\ref{eq:2}) is a hermitian one
by construction:
\begin{eqnarray}
K_{\rho\sigma}=K^*_{\sigma\rho}.
\end{eqnarray}
We will separate the contributions, associated with the
electron polarization:
\begin{eqnarray}
K_{\rho\sigma} &=& K^0_{\rho\sigma} + \frac{\alpha}{4\pi}
\biggl( K^1_{\rho\sigma} + K^{1*}_{\sigma\rho}\biggr), \\ \nonumber
K^0_{\rho\sigma} &=& B_{\rho\sigma} + \xi P^0_{\rho\sigma}, \qquad
K^1_{\rho\sigma}  =  T_{\rho\sigma} + \xi P^1_{\rho\sigma},
\end{eqnarray}
where $\xi$ is the degree of the initial electron polarization.
Quantities $B_{\rho\sigma}$ and $T_{\rho\sigma}$ correspond to the
case of unpolarized electron:
\begin{eqnarray}
B_{\rho\sigma} &=& B_g \tilde g_{\rho\sigma}
+ B_{11}\tilde p_{1\rho}\tilde p_{1\sigma}
+ B_{22}\tilde p_{2\rho}\tilde p_{2\sigma}, \\ \nonumber
B_g &=& \frac{1}{st}[(s+u)^2+(t+u)^2]-2m^2q^2\biggl(\frac{1}{s^2}
+ \frac{1}{t^2}\biggr), \\ \nonumber
B_{11} &=& \frac{4q^2}{st}-\frac{8m^2}{s^2},\qquad
B_{22}=\frac{4q^2}{st}-\frac{8m^2}{t^2}\, .
\end{eqnarray}
where the new variables
\begin{eqnarray}
\tilde g_{\rho\sigma} &=& g_{\rho\sigma} - \frac{q_\rho q_\sigma}{q^2}, \qquad
\tilde p_{1\rho} = p_{1,2}^{\rho}-\frac{p_{1,2}q}{q^2}q^\rho
\end{eqnarray}
provide an explicit fulfillment of gauge conditions:
$q_{\rho}K^{\rho\sigma} = 0$, $q_{\sigma}K^{\rho\sigma} = 0$.
Quantity $T_{\rho\sigma}$ has a rather cumbersome form,
it is given in~\cite{r1}.

For the case of the most general form for the electron polarization vector
\begin{eqnarray}
\Sigma u(p)\bar u(p)=(\hat p_1+m)(1-\xi\gamma_5 \hat a)
\end{eqnarray}
one obtains (see also \cite{KSh,ASh})
\begin{eqnarray}
P^0_{\rho\sigma} &=& 4m \biggl\{ (p_1q\rho\sigma)\frac{qa-2p_2a}{st}
+ (p_2q\rho\sigma)\biggl[ \frac{qa}{t^2}
+ \frac{p_2a}{t}\biggl(\frac{1}{s}-\frac{1}{t}\biggr) \biggr] \nonumber \\
&+& (qa\rho\sigma)\biggl[ \frac{q^2}{st} - \frac{1}{s}
- \frac{1}{t} - m^2\biggl(\frac{1}{s^2}
+ \frac{1}{t^2}\biggr)\biggr] \biggr\},
\end{eqnarray}
where we used the notation
\begin{eqnarray}
(abcd)\equiv {\mathrm{i}}\epsilon_{\alpha\beta\gamma\delta}
a^\alpha b^\beta c^\gamma d^\delta.
\end{eqnarray}
This object obeys the Shouten identity:
\begin{eqnarray}
(abcd)ef = (fbcd)ae + (afcd)be + (abfd)ce + (abcf)de.
\end{eqnarray}

In this paper we restrict ourselves by considering only the case
of longitudinally polarized fermion:
\begin{eqnarray}
\Sigma u(p_1)\bar u(p_1)=\hat p_1(1-\xi\gamma_5).
\end{eqnarray}
This is the most interesting case for physical applications.
In the Born approximation we obtain
\begin{eqnarray}
P_{\rho\sigma} &=& \xi \biggl[P^0_{\rho\sigma}
+ \frac{\alpha}{4\pi}P^1_{\rho\sigma}\biggr], \\ \nonumber
P^0_{\rho\sigma} &=& P^{0t}_{\rho\sigma} + P^{0s}_{\rho\sigma}
= \frac{2}{st} \biggl[ (u+t)(p_1q\rho\sigma) + (u+s)(p_2q\rho\sigma)\biggr].
\end{eqnarray}
Here and below the upper indexes $t$ and $s$ mean
the contributions of Feynman diagrams.
It is useful to present the explicit expressions for
$P^{0t,s}_{\rho\sigma}$:
\begin{eqnarray}
P^{0t}_{\rho\sigma} &=& \frac{1}{st}\biggl[ 4(p_1p_2q\sigma)(p_{1\rho}
+ p_{2\rho}) + 2(t-s)(p_1p_2\rho\sigma) + 2(s+u)(p_2q\rho\sigma)\biggr],
\nonumber \\
P^{0s}_{\rho\sigma} &=& \frac{1}{st}\biggl[
- 4(p_1p_2q\sigma)(p_{1\rho}+p_{2\rho})
+ 2(s-t)(p_1p_2\rho\sigma) + 2(s+t)(p_1q\rho\sigma)\biggr].
\end{eqnarray}
It is easy to check the following relations:
\begin{eqnarray}
q_{\rho}P^{0}_{\rho\sigma} = q_{\sigma}P^{0}_{\rho\sigma} = 0,\quad
(P^{0s,t}_{\sigma\rho})^* = P^{0s,t}_{\rho\sigma}\, , \quad
P^{0s,t}_{\rho\sigma}q_\rho=0, \quad
P^{0s,t}_{\rho\sigma}q_\sigma \ne 0.
\end{eqnarray}
Note now that we may consider in calculations only half of the full set of 8
Feynman diagrams in 1-loop level drawn in Fig.2, namely, the diagrams
(a),(b),(c),(d). Really the whole contribution may be obtained
knowing the values
of the contributions arising from Feynman diagrams Fig.2 using
the rearrangement operator:
\begin{eqnarray} \label{p1}
P^1_{\rho\sigma}=(1+\hat{H})(1-\hat{P})(P^{a,b}+P^{1c}+P^{1d})_{\rho\sigma}
+ P^{\mathrm{soft}}_{\rho\sigma},
\end{eqnarray}
where the operator $\hat{P}$ is defined as
\begin{eqnarray}
\hat{P} F(\rho,\sigma,p_1,p_2,q,s,t)=F(\rho,\sigma,p_2,p_1,-q,t,s),
\end{eqnarray}
and the hermitization operator $\hat{H}$ acts as:
\begin{eqnarray}
\hat{H}a_{\rho\sigma} = a^{*}_{\sigma\rho}\, .
\end{eqnarray}
Note that $\hat{P} P^{0s,t}_{\rho\sigma}=-P^{0t,s}_{\rho\sigma}$.
The last term in Eq.(\ref{p1}) describes the contribution due to
the emission of additional soft photon~\cite{r1}:
\begin{eqnarray}
P^{\mathrm{soft}}_{\rho\sigma} &=& P^0_{\rho\sigma}\delta^{\mathrm{soft}},
\\ \nonumber
\delta^{\mathrm{soft}} &=& - \frac{4\pi\alpha}{16\pi^3}\int
\frac{\dd^3 k}{\omega}(\frac{p_1}{p_1k}-\frac{p_2}{p_2k})^2=
\frac{\alpha}{\pi}\biggl[(L_u-1)
\ln\frac{m^2(\Delta \varepsilon)^2}{\lambda^2\varepsilon_1\varepsilon_2}
+ \frac{1}{2}L_u^2 \\ \nonumber
&-& \frac{1}{2}\ln^2\frac{\varepsilon_1}{\varepsilon_2}
- \frac{\pi^2}{3} + \Li\biggl(1+\frac{u}{4\varepsilon_1\varepsilon_2}\biggr)
\biggr], \qquad L_u = \ln\frac{-u}{m^2}\, .
\end{eqnarray}
Here $\Delta\varepsilon$ is the maximal energy of additional soft photon
escaping the detector; quantities $\varepsilon_{1,2} = p_{1,2}^0$ are
the energies of the initial and the final electron in the laboratory
reference frame (rest reference frame of the target).

Considering the contribution of Feynman diagrams Fig.2,a,b we may use
the result, given in the preprint of paper~\cite{r1}, namely
\begin{eqnarray}
(M^a_\sigma+M^b_\sigma)(-i(4\alpha\pi)^2)^{-1}=\frac{\alpha}{2\pi}\bar u(p_2)
\gamma_\sigma[m A_1(\hat e-\hat k_1\frac{p_1e}{p_1k_1})+A_2\hat k_1\hat e] u(p_1).
\end{eqnarray}
Note that this result may be reproduced using the loop integrals
list given in Appendix and the standard renormalization procedure.
We see that only structure in front of coefficient $A_2$ survives in
the limit
$m\to 0$. After simple algebra we obtain:
\begin{eqnarray}
P^{a,b}_{\rho\sigma}=2\frac{2L_t-1}{st}[2(p_1p_2q\sigma)p_{2\rho}+(u+s)((p_2q\rho\sigma)-
(p_1p_2\rho\sigma))].
\end{eqnarray}
The remaining Fig.2c,d contributions have a form:
\begin{eqnarray}
P^{1c}_{\rho\sigma}=\frac{1}{t}\int\frac{\dd^4 k}{\mathrm{i}\pi^2}\;
\frac{1}{a_0a_2a_q}\;\frac{1}{4}{\mathrm Tr}\,
\hat p_2\gamma_\lambda(\hat p_2-\hat k)\gamma_\sigma(\hat p_2-\hat q-\hat k)
\gamma_\lambda(\hat p_2-\hat q)\gamma_\mu\hat p_1\gamma_5\tilde O^{0}_{\rho\mu}
\end{eqnarray}
and
\begin{eqnarray}
P^{1d}_{\rho\sigma}=\int\frac{\dd^4 k}{\mathrm{i}\pi^2}\;
\frac{1}{a_0a_1a_2a_q}\;\frac{1}{4}{\mathrm Tr}\,
\hat p_2\gamma_\lambda(\hat p_2-\hat k)\gamma_\sigma(\hat p_2-\hat q-\hat k)
\gamma_\mu(\hat p_1-\hat k)\gamma_\lambda\hat p_1\gamma_5\tilde O^{0}_{\rho\mu}
\end{eqnarray}
where
\begin{eqnarray} \label{denoms}
a_0=k^2-\lambda^2,\quad
a_1=k^2-2p_1k,\quad a_2=k^2-2p_2k,\quad a_q=(p_2-q-k)^2-m^2,
\end{eqnarray}
and the matrix $\tilde O^{0}_{\rho\mu}$ differs from $O^{0}_{\rho\mu}$
(see Eq.(3)) by reversal order of gamma matrices.
Using the integrals given in Appendix one may perform the loop momenta
integration in right hand parts of expressions for $P^{1c}$, $P^{1d}$ and
obtain the total expression for the Compton tensor. Its explicit form
will be given below.

Now we will concentrate our attention on the terms containing the
infrared singularities. There are three sources of them. The first one is
the renormalization constant
\begin{eqnarray}
Z_1=1-\frac{\alpha}{2\pi}\biggl(\frac{1}{2}L_{\Lambda}
+2\ln\frac{\lambda}{m}+\frac{9}{4}\biggr), \quad
L_{\Lambda}=\ln\frac{\Lambda^2}{m^2},
\end{eqnarray}
which is needed to remove the ultraviolet divergence of the vertex
function, entering into $P^{1c}$. The next source is a part of the
box contribution $P^{1d}$, which comes from the terms from the numerator
which does not contain loop momenta. Really for the Feynman diagram Fig.2d
they are associated with the scalar integral,
\begin{eqnarray} \label{III}
I &=& \int\frac{\dd^4 k}{\mathrm{i}\pi^2}\;\frac{1}{a_0a_1a_2a_q}
= \frac{1}{tu}\biggl[
2L_u\ln\frac{m}{\lambda} - L_q^2 + 2L_tL_u - \frac{\pi^2}{6}
- 2\Li(1-\frac{q^2}{u}) \biggr], \\ \nonumber
L_q &=& \ln\frac{-q^2}{m^2},\quad L_t=\ln\frac{-t}{m^2}, \quad
\Li(z)=-\int\limits_{0}^{1}\frac{\dd x}{x}\ln(1-zx).
\end{eqnarray}
The third source is the emission of additional soft photons,
which was given above.
The infrared singularities are cancelled in the total sum.

Let us consider the contribution from one--loop corrections
(see Fig.2a,b,c,d)
\begin{eqnarray}
P^t_{\rho\sigma} = (P^{a,b}+P^{1c}+P^{1d})_{\rho\sigma}.
\end{eqnarray}

Extracting the leading logarithmic terms and infrared singularities,
we may present it as follows:
\begin{eqnarray}\label{0044}
P^t_{\rho\sigma} =P^{0t}_{\rho\sigma}\biggl[ - L_u^2
-4(L_u - 1)\ln\frac{m}{\lambda} + 3L_u \biggr]
+ R^t_{\rho\sigma}.
\end{eqnarray}

After hermitization and rearrangement operations and
adding of the soft photon contribution we arrive to the result
\begin{eqnarray}
P_{\rho\sigma} =P^{0}_{\rho\sigma}\biggl\{ 1 + \frac{\alpha}{\pi}
\biggl[ (L_u-1)\ln\frac{(\Delta\varepsilon)^2}{\varepsilon_1\varepsilon_2}
+ \frac{3}{2}L_u - \frac{1}{2}\ln^2\frac{\varepsilon_2}{\varepsilon_1}
- \frac{\pi^2}{3} + \Li(\cos^2\frac{\theta}{2}) \biggr] \biggr\}
+ \frac{\alpha}{4\pi}R_{\rho\sigma}.
\end{eqnarray}
Quantities $R^t_{\rho\sigma}$ and $R_{\rho\sigma}$ collect
non--leading terms. They are free from infrared singularities.

Tensor $R^t_{\rho\sigma}$ can be presented in the form
\begin{eqnarray}
R^t_{\rho\sigma} &=& A(2q\sigma\rho) + B(1q\sigma\rho)
+ C(12q\sigma)p_{1\rho} + D(12q\sigma)p_{2\rho}
+ E(12q\sigma)q_{\rho} + F(12\sigma\rho).
\end{eqnarray}
The coefficients $A-F$ have a rather cumbersome form, we are not going
to present them here. Note only that they obey the condition
\begin{eqnarray}
Cp_1q + Dp_2q + Eq^2 - F = 0,
\end{eqnarray}
because of gauge invariance in respect to index $\rho$.

The rearrangement operation gives
\begin{eqnarray}
(1-\hat{P})R^t_{\rho\sigma} &=&
(A+\tilde{B})(2q\sigma\rho) + (B+\tilde{A})(1q\sigma\rho)
+ (C-\tilde{D})(12q\sigma)p_{1\rho}
\nonumber \\ &+& (D-\tilde{C})(12q\sigma)p_{2\rho}
+ (E+\tilde{E})(12q\sigma)q_{\rho} + (F+\tilde{F})(12\sigma\rho)
\\ \nonumber
&\equiv& A_1(1q\sigma\rho) + A_2(2q\sigma\rho)
+ B_1(12q\sigma)p_{1\rho} + B_2(12q\sigma)p_{2\rho}
\\ \nonumber
&+& C_1(12q\sigma)q_{\rho} + F_1(12\sigma\rho).
\end{eqnarray}
Tests of gauge invariance gives an important check of our calculations:
\begin{eqnarray}
q^{\rho}(1-\hat{P})R_{\rho\sigma} &=&
B_1(12q\sigma)p_1q + B_2(12q\sigma)p_2q
+ C_1(12q\sigma)q^2  \nonumber \\
&+& F_1(12\sigma q) = 0,\qquad
q^{\sigma}(1-\hat{P})R_{\rho\sigma} = F_1(12q\rho) = 0.
\end{eqnarray}
The above conditions yield
\begin{eqnarray}
F_1 = 0, \qquad
C_1 = - B_1\frac{p_1q}{q^2} - B_2\frac{p_2q}{q^2}\, , \nonumber \\
B_1p_{1\rho} + B_2p_{2\rho} + C_1q_{\rho}
= B_1\tilde{p}_{1\rho} + B_2\tilde{p}_{2\rho}\, , \\ \nonumber
\tilde{p}_{1\rho} = p_{1\rho} - \frac{p_1q}{q^2}q_{\rho}\, , \quad
\tilde{p}_{2\rho} = p_{2\rho} - \frac{p_2q}{q^2}q_{\rho}\, .
\end{eqnarray}
By straightforward calculations we checked these relations.

The hermitization gives
\begin{eqnarray}
R_{\rho\sigma} &=& (1+\hat{H})(1-\hat{P})R^t_{\rho\sigma}
= (A_1 + A_1^*)(1q\sigma\rho) + (A_2 + A_2^*)(2q\sigma\rho)
+ (12q\sigma) [B_1\tilde{p}_{1\rho} + B_2\tilde{p}_{2\rho}] \nonumber \\
&-& (12q\rho)[B_1^*\tilde{p}_{1\sigma} + B_2^*\tilde{p}_{2\sigma}],
\end{eqnarray}
where
\begin{eqnarray}\label{A1B1}
A_1 &=& \frac{2}{st}\biggl[
\frac{2u(2s-u)}{a}L_{qu} + \frac{4us}{a}\biggl(\frac{u}{a}L_{qu}-1\biggr)
+ \frac{ub}{c} + \frac{2u^2+us-s^2}{c}L_{sq} + \frac{usb}{c^2}L_{sq}
\nonumber \\
&-& 2c\zeta(2) - 2cL_{tu} + (2s-c)L_{qu} - \frac{uc}{s}G
+ \biggl(\frac{ub}{t}+c\biggr)\tilde{G} + 5c - 2s \biggr],
\\ \nonumber
B_1 &=& \frac{2}{st}\biggl[ \frac{8u}{a}\biggl( 1
- \biggl(\frac{u}{a}+1\biggr)L_{qu} \biggr) + \frac{6t}{b}L_{qt}
+ \frac{2(u^2-2s^2-su)}{cu}L_{sq}
\\ \nonumber
&+& \frac{2b}{c}\biggl(1+\frac{s}{c}L_{sq}\biggr)
+ \frac{2}{s}(2c-s)L_{tu} + \biggl( - 2 - \frac{4c^2}{st} - \frac{12b}{t}
- \frac{4s^2}{ut}\biggr)L_{qu}
\\ \nonumber
&+& \frac{4b^2}{tu}L_{su} + \biggl(-2+\frac{2uc}{s^2}-\frac{2t}{s}\biggr)G
+ \biggl(\frac{2b}{t}+\frac{2b^2}{t^2}\biggr)\tilde{G} + 6 \biggr],
\\ \nonumber
G &=& (L_q - L_u)(L_q + L_u - 2L_t) - \frac{\pi^2}{3}
- 2\Li\biggl(1-\frac{q^2}{u}\biggr) + 2\Li\biggl(1-\frac{t}{q^2}\biggr),
\\ \nonumber
A_2 &=& (s\leftrightarrow t) A_1, \qquad
B_2 = - (s\leftrightarrow t) B_1, \qquad
\tilde{G} = (s\leftrightarrow t) G.
\end{eqnarray}
Note that the above expressions are free from kinematical singularities.
Really, in the limits $a\to 0$, $b\to 0$ and $c\to 0$ the
quantities are finite. The symmetry between $A_1$, $B_1$ and $A_2$, $B_2$
is because of the initial symmetry between $p_1$ and $p_2$ in
the traces.


Thus we calculated the part of the leptonic tensor, proportional
to the initial longitudinal polarization. This tensor describes
Compton scattering with one off--shell photon, which is connected
with a certain target.

The calculation allows to obtain the correction coming
from one--loop effects to quantities observable in different
polarization experiments.
Let us consider for definiteness the task Of calculation  of $\alpha^2$
order radiative
correction  in polarized deep inelastic scattering.
The results for the lowest order QED correction for nucleon and nuclear
targets can be found
in refs.\cite{KSh,ASh}.
Both the Born cross--section ($\sigma_{\mathrm{Born}}$) and
the cross--section at the level of radiative corrections
($\sigma_{\mathrm{RC}}$) can be split into unpolarized and polarized
parts
\begin{equation}
\sigma_{\mathrm{Born,RC}}=\sigma_{\mathrm{Born,RC}}^{\mathrm{unp}}
+ \xi_b\xi_t\sigma_{\mathrm{Born,RC}}^{\mathrm{pol}},
\end{equation}
where $\xi_b$ and $\xi_t$ are polarization degrees of beam and target.
The correction to asymmetry ($A=\sigma^{\mathrm{pol}}/\sigma^{\mathrm{unp}}$):
\begin{equation}
\Delta
A={\sigma^{\mathrm{pol}}_{\mathrm{RC}}\sigma^{\mathrm{unp}}_{\mathrm{Born}}
-\sigma^{\mathrm{unp}}_{\mathrm{RC}}\sigma^{\mathrm{pol}}_{\mathrm{Born}}
\over
\sigma^{\mathrm{unp}}_{\mathrm{Born}}(\sigma^{\mathrm{unp}}_{\mathrm{Born}}
+\sigma^{\mathrm{unp}}_{\mathrm{RC}})}
\label{asym}
\end{equation}
is usually not large
because of mutual cancellation of large factorizing terms
in eqn.(\ref{asym}). It is clear that in such cases when relatively
small correction is obtained as a difference of two large terms the
radiative correction cross--section has to be calculated with the most
possible accuracy, and special attention has to be paid to
non--factorizing terms like (\ref{A1B1}).

Now the new methods of experimental data processing, where
experimental information about spin observables is extracted directly
from polarized part of cross--section (difference of observed cross
sections with opposite spin configurations)~\cite{Gagu} is actively
developed. It makes new requirements for accuracy of radiative
correction calculation. We note that
there is no any cancellation of leading contributions in
this case, and factorizing terms in (\ref{0044}) give the basic contribution.

The kinematical regions with very high $y$ ($y\sim 0.9$) can be
reachable in the current polarization experiments on DIS
\cite{SMC,HERMES}. In this region radiative correction to
cross--section is comparable or larger of Born cross--section.
Basically it is
originated by contributions of radiative tails from elastic and
quasielastic peaks. This calculation firstly allows to obtain the
contribution of these tails with taking into account loop effects in the
nest--to--leading approximation.

There is one particular interesting phenomenon. Note, that
$P^{(1)}_{\rho\sigma}$ contains not only the imaginary part, but
also a certain real part, which comes from the imaginary parts of
$A_1$ and $B_1$. The conversion of this real part of
$P^{(1)}_{\rho\sigma}$ with the ordinary symmetrical part of the
hadronic tensor will give rise to one--spin azimuthal asymmetry
for the final electron~\cite{onesp}. The asymmetry is proportional
to the degree of polarization of the initial electron. It is small
because of the extra power of $\alpha_{\mathrm{QED}}$ and the absence
of large logarithms.

\subsection*{Acknowledgments}
The authors are indebted to M.~Galynski, T.~Shishkina and
N. Shumeiko for fruitful discussions and to B.~Shaikhatdenov
for help.
Two of us (A.A., E.K.) are grateful to the INTAS foundation,
grant 93--1867 ext.

\section*{Appendix}
\setcounter{equation}{0}
\renewcommand{\theequation}{A.\arabic{equation}}

Here we put the list of relevant one--loop integrals calculated in the
approximation $s\sim -t\sim-u\sim -q^2\gg m^2$.
We use the notation:
\begin{eqnarray}
(I,I_\mu,I_{\mu\nu}) &=& \int\frac{\dd^4 k}{\mathrm{i}\pi^2}
\frac{(1,k_\mu,k_\mu k_\nu)}{a_0a_1a_2a_q}\, ,
\qquad
(i,i_\mu,i_{\mu\nu})=\int\frac{\dd^4 k}{\mathrm{i}\pi^2}
\frac{(1,k_\mu,k_\mu k_\nu)}{a_1a_2a_q}\, ,
\nonumber \\
(j,j_\mu,j_{\mu\nu}) &=& \int\frac{\dd^4 k}{\mathrm{i}\pi^2}
\frac{(1,k_\mu,k_\mu k_\nu)}{a_0a_2a_q}\, ,
\quad
(n,n_\mu,n_{\mu\nu})=\int\frac{\dd^4 k}{\mathrm{i}\pi^2}
\frac{(1,k_\mu,k_\mu k_\nu)}{a_0a_1a_q}\, ,
\nonumber  \\
(z,z_\mu,z_{\mu\nu}) &=& \int\frac{\dd^4 k}{\mathrm{i}\pi^2}
\frac{(1,k_\mu,k_\mu k_\nu)}{a_0a_1a_2}\, .
\end{eqnarray}
Using the explicit expression for the denominators~(\ref{denoms}),
we obtain a decomposition for vector--type integrals:
\begin{eqnarray}
I_\mu &=& I_1p_{1\mu} + I_2p_{2\mu} + I_{q}q_{\mu}, \quad
i_\mu = i_1p_{1\mu} + i_2p_{2\mu} + i_{q}q_{\mu}, \quad
n_\mu = n_1p_{1\mu}+n_2(p_{2\mu}-q_{\mu}),\nonumber \\
j_\mu &=& j_2p_{2\mu}+j_{q}q_{\mu}, \quad
z_\mu = z_1(p_{1\mu} + p_{2\mu}),
\end{eqnarray}
In the same way we get for tensor--type integrals:
\begin{eqnarray}
I_{\mu\nu} &=& I_g g_{\mu\nu} + I_{11}p_{1\mu}p_{1\nu}
+ I_{22}p_{2\mu}p_{2\nu} + I_{qq}q_{\mu}q_{\nu}
+ I_{12}(p_{1\mu}p_{2\nu}+p_{2\mu}p_{1\nu}) \nonumber \\
&+& I_{1q}(p_{1\mu}q_{\nu}+q_{\mu}p_{1\nu})
+ I_{2q}(p_{2\mu}q_{\nu}+q_{\mu}p_{2\nu}), \nonumber \\
i_{\mu\nu} &=& i_g g_{\mu\nu} + i_{11}p_{1\mu}p_{1\nu}
+ i_{22}p_{2\mu}p_{2\nu} + i_{qq}q_{\mu}q_{\nu}
+ i_{12}(p_{1\mu}p_{2\nu}+p_{2\mu}p_{1\nu})
 \nonumber \\
&+& i_{1q}(p_{1\mu}q_{\nu}+q_{\mu}p_{1\nu})
+ i_{2q}(p_{2\mu}q_{\nu}+q_{\mu}p_{2\nu}),
 \nonumber \\
j_{\mu\nu} &=& j_g g_{\mu\nu} + j_{22}p_{2\mu}p_{2\nu}
+ j_{qq}q_{\mu}q_{\nu} + j_{2q}(p_{2\mu}q_{\nu}+q_{\mu}p_{2\nu}),
 \nonumber \\
n_{\mu\nu} &=& n_g g_{\mu\nu} + n_{11}p_{1\mu}p_{1\nu}
+ n_{22}(p_{2\mu}p_{2\nu} - q_{\mu}p_{2\nu} - p_{2\mu}q_{\nu}
+ q_{\mu}q_{\nu})
\nonumber \\
&-& n_{1q}[p_{1\mu}(p_{2\nu}-q_{\nu}) + p_{1\nu}(p_{2\mu}-q_{\mu})],
\nonumber \\
z_{\mu\nu}&=&z_g g_{\mu\nu}+z_{11}(p_{1\mu}p_{1\nu}+p_{2\mu}p_{2\nu})+
z_{12}(p_{2\mu}p_{1\nu}+p_{1\mu}p_{2\nu}),
\end{eqnarray}
The quantities, entering into vector and tensor integrals, are:
\begin{eqnarray}
I_1&=&\frac{1}{d}\biggl[(ut-sq^2)i+b^2j+t(s-u)n-ubY\biggr], \nonumber \\
I_2&=&\frac{1}{d}\biggl[(us-tq^2)i+(ts-uq^2)j+tcn+ucY\biggr], \nonumber \\
I_q&=&\frac{1}{d}\biggl[u(t-s)i+ubj-utn-u^2Y\biggr],\\ \nonumber
Y &=& z-tI, \quad d=2stu, \quad b=u+s, \quad  a=s+t, \quad c=u+t.
\end{eqnarray}
For coefficients in tensor structures of $I_{\mu\nu}$ we have
\begin{eqnarray}
I_g&=&\frac{1}{2}(i+tI_q), \qquad
I_{11}=\frac{1}{d}\biggl[ b^2(i+tI_q)+(ut-sq^2)i_1+t(s-u)n_1-ub(z_1-tI_1)
\biggr], \nonumber \\
I_{22}&=&\frac{1}{d}\biggl[ c^2(i+tI_q) + (us-tq^2)i_2 + (ts-uq^2)j_2
+ tcn_2 + uc(z_1-tI_2) \biggr], \nonumber \\
I_{12}&=&\frac{1}{d}\biggl[ (st-uq^2)(i+tI_q) + (us-tq^2)i_1 + tcn_1
+ uc(z_1-tI_1) \biggr], \nonumber \\
I_{1q}&=&\frac{1}{d}\biggl[ bu(i+2tI_q) + (ut-sq^2)i_q + b^2j_q
+ t(u-s)n_2 \biggr], \nonumber \\
I_{2q}&=&\frac{1}{d}\biggl[ - uc(i+2tI_q) + (us-tq^2)i_q
+ (ts-uq^2)j_q - tcn_2 \biggr], \nonumber \\
I_{qq}&=&\frac{1}{d}\biggl[ u^2(i+2tI_q) + u(t-s)i_q + tun_2
+ ubj_q \biggr].
\end{eqnarray}
The $i$--vector--type integrals are read
\begin{eqnarray}
i_1 &=& \frac{1}{a^2}\biggl[ q^2ai+(q^2+u)L_u-2q^2L_q+2a\biggr],\qquad
i_2=i-i_1, \nonumber \\
i_q &=& \frac{1}{a^2}\biggl[uai+2uL_u-(q^2+u)L_q+2a\biggr].
\end{eqnarray}
The tensor--type integrals are
\begin{eqnarray}
i_g&=&\frac{1}{4}L_{\Lambda}+\frac{3}{8}+\frac{1}{4a}(uL_u-q^2L_q),
\nonumber \\
i_{11}&=&\frac{1}{a^3}\biggl[ (q^2)^2ai
+ \frac{1}{2}(3(q^2)^2+4q^2u-u^2)L_u - 3(q^2)^2L_q + a(4q^2-u) \biggr],
\nonumber \\
i_{22}&=&\frac{1}{a^3}\biggl[ u^2ai
+ \frac{1}{2}(-(q^2)^2+4q^2u+3u^2)L_u + q^2(q^2-4u)L_q + 3ua \biggr],
\nonumber \\
i_{qq}&=&\frac{1}{a^3}\biggl[ u^2ai + 3u^2L_u
+ \frac{1}{2}((q^2)^2-4q^2u-3u^2)L_q + a(4u-q^2) \biggr],
\nonumber \\
i_{12}&=&\frac{1}{a^3}\biggl[ - uq^2ai
- \frac{1}{2}((q^2)^2+4q^2u+u^2)L_u + q^2(q^2+2u)L_q - a(2q^2+u) \biggr],
\nonumber \\
i_{1q}&=&\frac{1}{a^3}\biggl[ uq^2ai + \frac{u}{2}(5q^2+u)L_u
- \frac{q^2}{2}(q^2+5u)L_q + \frac{3a}{2}(q^2+u) \biggr],
\nonumber \\
i_{2q}&=&\frac{1}{a^3}\biggl[ - u^2ai - \frac{u}{2}(q^2+5u)L_u
+ \frac{1}{2}(-(q^2)^2+5uq^2+2u^2)L_q + \frac{a}{2}(q^2-7u) \biggr].
\end{eqnarray}
The remaining vectors coefficients are:
\begin{eqnarray}
j_2&=&\frac{2q^2}{b^2}L_q - \frac{q^2+t}{b^2}L_t - \frac{t}{b}j, \qquad
j_q=\frac{1}{b}(L_t-L_q), \nonumber \\
n_1&=&\frac{1}{t}(tn-2L_t+2), \qquad
n_2=\frac{1}{t}(L_t-2), \qquad
z_1=\frac{1}{u}L_u.
\end{eqnarray}
Remaining tensor coefficients are:
\begin{eqnarray}
j_g&=&\frac{1}{4}L_{\Lambda}+\frac{3}{8}+\frac{1}{4b}(tL_t-q^2L_q), \quad
j_{qq}=\frac{1}{2b}(L_q-L_t),
\nonumber \\
j_{2q}&=&\frac{1}{2b^2}\biggl[(q^2-2t)(L_t-L_q)-b\biggr],
\nonumber \\
j_{22}&=&\frac{1}{b^3}\biggl[ \frac{1}{2}(4q^2t+3t^2-(q^2)^2)L_t
+ q^2(q^2-4t)L_q + \frac{b}{2}(q^2+t) + t^2bj \biggr],
\nonumber \\
n_g&=&\frac{1}{4}L_{\Lambda}-\frac{1}{4}L_t+\frac{3}{8}, \qquad
n_{11}=n+\frac{1}{t}(-3L_t+\frac{9}{2}),
\nonumber \\
n_{22}&=&\frac{1}{2t}(L_t-2), \qquad
n_{1q}=\frac{1}{2t}(-L_t+3),
\nonumber \\
z_g &=& \frac{1}{4}(L_{\Lambda}-L_u+\frac{3}{2}), \qquad
z_{11}=\frac{1}{2u}, \qquad
z_{12}=\frac{1}{2u}(L_u-1).
\end{eqnarray}
We use the notation
\begin{eqnarray}
L_{\Lambda}=\ln\frac{\Lambda^2}{m^2},\qquad L_t=\ln\frac{-t}{m^2},
\qquad L_u=\ln\frac{-u}{m^2}\, .
\end{eqnarray}
The ultraviolet cut--off momentum parameter $\Lambda$ will be eliminated
from the final answer after the renormalization procedure. Really, accounting
the renormalization constant
\begin{eqnarray}
Z_1 = - \frac{\alpha}{2\pi}\biggl( \frac{1}{2}L_{\Lambda}
+ 2\ln\frac{\lambda}{m} + \frac{9}{4} \biggr)
\end{eqnarray}
for lepton Green functions, we get
\begin{eqnarray}
L_{\Lambda} \to -4\ln\frac{\lambda}{m}-\frac{9}{2} \, ,
\end{eqnarray}
where $\lambda$ is a fictitious photon mass, $\lambda\ll m$.
We put below the scalar integrals:
\begin{eqnarray}
i&=&\frac{1}{2a}(L_q^2-L_u^2), \qquad
z=\frac{1}{2u}\biggl[ 4\ln\frac{m}{\lambda}L_u + L_u^2
- \frac{\pi^2}{3}\biggr], \nonumber \\
n&=&\frac{1}{2t}\biggl[L_t^2+\frac{2\pi^2}{3}\biggr], \qquad
j=\frac{1}{b}\biggl[L_q(L_q-L_t) + \frac{1}{2}(L_q-L_t)^2
+ 2\Li\biggl(1-\frac{t}{q^2}\biggr)\biggr].
\end{eqnarray}
Note that the quantity $Y=z-tI$ (the scalar integral
$I$ is given in Eq.(\ref{III}))
is free from infrared singularities:
\begin{eqnarray}
Y=\frac{1}{u}\biggl[ \frac{1}{2}L_u^2 + L_q^2 - 2L_uL_t
+ 2\Li\biggl(1-\frac{q^2}{u}\biggr) \biggr].
\end{eqnarray}

\end{document}